\documentclass[preprint,showpacs,preprintnumbers,amsmath,amssymb,nofootinbib,showkeys,aps]{revtex4}
\pdfoutput=1
\usepackage{epsfig}

\usepackage{graphicx}
\usepackage{dcolumn}
\usepackage{bm}
\usepackage{amsmath}
\usepackage{amssymb}
\usepackage{latexsym}
\usepackage{color}
\usepackage{hyperref}
\usepackage{mathrsfs}

\newcommand{\be}{\begin{equation}}
\newcommand{\ee}{\end{equation}}
\newcommand{\bea}{\begin{eqnarray}}
\newcommand{\eea}{\end{eqnarray}}
\newcommand{\texpdf}{\texorpdfstring}
\newcommand{\like}{\mathscr{L}}
\begin{document}


\title{Determination of the Kinematic Parameters from SNe Ia and Cosmic Chronometers}

\author{D. Benndorf$^{1}$} \email{douglas.benndorf@unesp.br}
\author{J. F. Jesus$^{1,2}$}\email{jf.jesus@unesp.br}
\author{S. H. Pereira$^{1}$} \email{s.pereira@unesp.br}

\affiliation{$^1$Departamento de F\'isica, Faculdade de Engenharia de Guaratinguet\'a, Universidade Estadual Paulista (UNESP),  Av. Dr. Ariberto Pereira da Cunha 333, 12516-410, Guaratinguet\'a, SP, Brazil. \\
\\$^2$Instituto de Ciências e Engenharia, Universidade Estadual Paulista (UNESP),  R. Geraldo Alckmin, 519, 18409-010, Itapeva, SP, Brazil.
}


\def\zt{\mbox{$z_t$}}

\begin{abstract}

In this work, by assuming a spatially flat Universe, we have tested 8 kinematic parametrization models with $H(z)$ data from Cosmic Chronometers and SNe Ia from Pantheon compilation. Our aim is obtain the current values for the Hubble constant ($H_0$), deceleration parameter ($q_0$), jerk ($j_0$) and snap ($s_0$) parameters independently from a dynamical model. By using a Bayesian model comparison, three models are favoured: a model with the deceleration parameter ($q$) linearly dependent on the redshift, $q$ linearly dependent on the scale factor and a model with a constant jerk. The model with constant jerk is slightly favoured by this analysis, furnishing $H_0=68.8^{+3.7}_{-3.6}$ km/s/Mpc, $q_0=-0.58\pm0.13$, $j_0=1.15^{+0.56}_{-0.53}$ and $s_0=-0.25^{+0.40}_{-0.30}$. The other models are compatible with the constant jerk model, except for the snap parameter, where we have found $s_0=4.0^{+3.4}_{-3.0}$ for the model with $q$ linearly dependent on the scale factor\footnote{{All uncertainties in the Abstract correspond to 95\% c.l.}}.
\end{abstract}

\maketitle



\section{Introduction}

Observations of differential
age of distant galaxies through Hubble parameter ($H(z)$) data \cite{MaganaEtAl17} and Supernovae Type Ia (SNe Ia) \cite{pantheon} compilation over the past few decades show that the universe is entering an accelerated expansion. Although the standard $\Lambda$CDM model fits quite well the observational data, alternative models with different energetic contents have also been invoked in order to deal with some problems suffered by the standard model \cite{Bull2016}. In some of these models a new dynamical component enters the equations to mimic the dark energy and dark matter effects, or even the interaction among them \cite{GongBo,marttens2020,SF, Majerotto2009,Valiviita2010,Chimento2010,Cai2010,Sun2012,Pourtsidou2013,Salvatelli2014,Li2014,Skordis2015,Jimenez2016,Valent2020}.

Other possibilities are the cosmographic (or kinematic) models \cite{kine1,kine2,kine3,kine4,kine5,kine6,kine7}, where it is not assumed any dynamic energetic content and we seek for direct measures of expansion through its kinematic parameters, such as the Hubble parameter, deceleration parameter, jerk and snap parameters, etc. In these methods no relationship between mass-energy and geometry is assumed, solely that the geometry of the Universe is pseudo-Riemmanian. The advantage of use cosmographic modeling is that it has less bias, since it seeks for a direct measure of expansion parameters from the data, with fewer assumptions than in dynamic modeling, which means that the parameters obtained via cosmography must be more trustworthy to reality. One disadvantage is that, since a relationship between energy content and geometry is not assumed, the measurements obtained from the data tend to have greater statistical uncertainties. Different types of dark energy models in the framework of the cosmographic approach, with emphasis
on the running vacuum models has been studied recently in \cite{rezaei2021} and cosmographic functions up to the fourth derivative of the scale factor using the non-parametric method of Gaussian
Processes was done in \cite{velasques2021}. 

One way to carry out the cosmographic modeling is to parameterize the deceleration, jerk and snap parameters ($q$, $j$ and $s$, respectively), or even higher orders if desired (crackle, pop etc). From this parameterization we can obtain other parameters. An study for the redshift drift in terms of the present day Hubble, deceleration, jerk, snap and other parameters has been done in \cite{lobo2020} and a joint analysis using BAO, Hubble data and Pantheon compilation of Supernovae type Ia for a parametrization with a constant jerk parameter was done by \cite{hassan2020} obtaining $j_0 = 1.038^{+0.061}_{-0.023}$. {In \cite{valent19} two different  expansions for $q$ as a function of the redshift were carried out, providing the constraints $q_0 = -0.43^{+0.04}_{-0.07}$ and $j_0=1.5^{+1.0}_{-0.7}$ for a polynomial expansion. An explicit reconstruction of the jerk parameter in a non-parametric way from model independent observational
data was done recently by \cite{Mukh}}.

Parameterization can be done in several ways, from the most basic linear dependence on redshift, to more elaborate, such as Padé expansions \cite{pade}, Chebyshev polynomials \cite{cheb} and logarithmic polynomials \cite{log}. Alternative parameterizations are necessary due to the non-convergence of the Taylor series for redshift $z\ge 1$, which means that even the truncation of the expansion implies a poor modeling for higher redshifts. One of the simplest ways to get around this problem is to redefine the standard redshift relation  $ 1+z\equiv {\lambda_0}/{\lambda_e}$ as $1+y={\lambda_e}/{\lambda_0}$, so that we have now a new $y$-redshift parameter \cite{yred}. The advantage of use $y$-redshift is that its convergence interval [0, 1] corresponds to the whole period from the origin of the universe to the present moment, while it diverges just for $y \to -1$, corresponding to the distant future (the inverse occurs with the $z$-redshift). However, as the observational data are all in the past, the $y$-redshift is ideal for handling with it.

The $y$-redshift can be obtained from $z$-redshift via the relation
\begin{equation}
    y=\frac{z}{1+z}
\end{equation}
which means that all observational data already collected for $z$-redshift can be directly used in analysis involving $y$-redshift. Since 
\begin{equation}
    y=1-\frac{1}{1+z} = 1-a(z),
\end{equation}
we can understand the Maclaurin series expansions in $y$ as Taylor-like series expansions in the scale factor around $a=1$ (today). 

{Although, in terms of convergence for $z>1$, $y$-redshift performs better than $z$-redshift, $z$-redshift is still better for $z<1$. This is because $y$-redshift implies greater uncertainties, as shown by \cite{colgain} and in our results in Table \ref{results}. The factor $1-a$ is a parameter smaller than $z$-redshift in the same redshift range, so one needs more data to constrain the parameters in the same way. That is, the choice of $y$-redshift or $z$-redshift is a trade-off.}

In this work we obtain and compare the present cosmological parameters $q_0$, $j_0$ and $s_0$ through the linear parameterizations of $q$, $j$ and $s$ as a function of $z$ or $y$, in the general forms $u(x)=u_0$ and $u(x)=u_0+u_1x$, where $u=(q, j, s)$ and $x=(z, y)$. {We choose eight different parameterizations, which differ from other works in the sense that they cover most of the kinematic quantities, namely, 2nd, 3rd and
4th scale factor derivative and also cover the most popular parameterizations in the literature, namely constant and linear in the redshift $z$ and in the $y$-redshift. Additionally, in order to make a model comparison among all parameterizations, we have calculated the Bayesian Information Criterion (BIC), which allows to choose the three most favoured model. Also, we have worked with the
most recent SNe Ia and $H(z)$ data available.}

The paper is organized as follows: In Section II we present the main equations involving cosmological kinematic parameters. In Section III we present the kinematic parameterization models used. In Section IV we present the analysis and results for the models and finish with Conclusions on Section V.

\section{Equations for kinematic parameters}

The cosmological kinematic parameters are given by:
\begin{align}
    H&= \frac{\dot{a}}{a}\label{Hdef}\\
    q&=-\frac{\ddot{a}}{aH^2}\label{qdef}\\
    j&= \frac{\dddot{a}}{aH^3}\label{jdef}\\
    s&= \frac{\ddddot{a}}{aH^4}
\end{align}
for Hubble, deceleration, jerk and snap parameters, respectively, where $a(t)$ is the scale factor of Friedmann-Robertson-Walker metric, $\dot{a} \equiv da/dt$ and $H = \dot{a}/a$. The negative sign on the deceleration parameter $q$ is due to historical reasons. 

The kinematic parameters are often written in terms of the redshift $z$ through the definition $1+z \equiv 1/a$, from which follows the  relation $\frac{d}{dt}=-H(1+z)\frac{d}{dz}$, where we have used the present day scale factor equal unit, $a_0 =1$.
Thus, the deceleration parameter \eqref{qdef} can be written as:
\begin{align}
q&=-\frac{1}{aH^2}\frac{d(\dot{a})}{dt}=(1+z)\frac{H'}{H}-1\,,
\label{qHz}
\end{align}
where $H'\equiv dH/dz$. From this relation, $H(z)$ can be obtained from a specific $q(z)$ parametrization as:
\begin{align}
    \frac{1}{H}\frac{dH}{dz}=\frac{1+q(z)}{1+z}\Rightarrow\int_{H_0}^H\frac{dH}{H}=\int_0^z\frac{1+q(z)}{1+z}dz
\end{align}
from which follows
\begin{align}
     H=H_0\exp\left[\int_0^z\frac{1+q(z)}{1+z}dz\right]
    \label{Hfromq}
\end{align}

The jerk parameter \eqref{jdef} can be written as:
\begin{align}
j=\frac{1}{aH^3}\frac{d}{dt}(\ddot{a})=1-2(1+z)\frac{H'}{H}+(1+z)^2\frac{H'^2}{H^2}+(1+z)^2\frac{H''}{H}\,,\label{eq10}
\end{align}
and for a specific $j(z)$ parametrization, we have the equation for $H(z)$:
\be
H''(z)-\frac{2H'(z)}{1+z}+\frac{H'(z)^2}{H(z)}-\frac{H(z)(j(z)-1)}{(1+z)^2}=0\,.
\label{odeHj}
\ee
If one does the substitution $f(z)\equiv H(z)^2$, \eqref{odeHj} reads:
\be
(1+z)^2f''(z)-2(1+z)f'(z)+2\left[1-j(z)\right]f(z)=0
\label{odefzgeral}
\ee
which is a second order linear differential equation on $f(z)$.

For higher order parametrizations, it is more convenient to work with a system of first order differential equations.
In order to do that, let us define a generic kinematic parameter of order $n\geq2$, $\ell_n$:
\be
\ell_n\equiv\frac{a^{(n)}}{aH^n}
\label{lndef}
\ee
where $a^{(n)}\equiv d^n a/dt^n$.
Naturally, a higher order parameter $\ell_{n+1}$ will be:
\be
\ell_{n+1}\equiv\frac{a^{(n+1)}}{aH^{n+1}}\,,
\ee
in such a way that:
\begin{align}
a^{(n)}&=aH^n\ell_n\\
a^{(n+1)}&=\frac{d}{dt}\left(a^{(n)}\right)=-H(1+z)\frac{d}{dz}\left(aH^n\ell_n\right)=aH^{n+1}\ell_{n+1}
\end{align}
and
\begin{align}
\frac{H^{n+1}\ell_{n+1}}{1+z}&=-H(1+z)\left[\frac{nH^{n-1}H'\ell_n}{1+z}-\frac{H^n\ell_n}{(1+z)^2}+\frac{H^n\ell_n'}{1+z}\right]\,.\label{eq17}
\end{align}
According to \eqref{qHz}:
\be
H'= \frac{H}{1+z}(1+q)\,,
\label{Hlinhaq}
\ee
so, using \eqref{eq17}, we may write:
\be
\ell_{n+1}=-n(1+q)\ell_n+\ell_n-\ell_n'(1+z)\label{odeln}
\ee
from which we can obtain $\ell'_n$ for a given $q(z)$ parameterization.

 For $n=2$, with $\ell_2=-q$, $\ell_3=j$, \eqref{odeln} reads\footnote{The deceleration parameter $q$ is the only kinematic parameter which changes sign with respect to the definition \eqref{lndef}.}:
\be
(1+z)q'=j(z)-q(1+2q),
\label{odejz}
\ee
so, from  \eqref{Hlinhaq} and \eqref{odejz}, one may solve a model with a given jerk parametrization from the system:
\be
\left\{\begin{array}{ll}
     (1+z)H'&= (1+q)H \\
     (1+z)q'&=j-2q^2-q
\end{array}\right.
\label{sysjerk}
\ee

For $n=3$, with  $\ell_3=j$, $\ell_4=s$, \eqref{odeln} reads:
\be
(1+z)j'=-s(z)-(2+3q)j
\ee
So, one can solve a model with a given $s(z)$ parametrization from the system:
\be
\left\{\begin{array}{ll}
     (1+z)H'&= (1+q)H \\
     (1+z)q'&=j-2q^2-q\\
     (1+z)j'&=-s(z)-(2+3q)j
\end{array}\right.
\label{sysqH}
\ee
The process can be repeated for higher order parameters.

\section{Kinematic Parametrizations}

The parametrizations we will use on this paper are summarized in the Table I below.

\begin{table}[ht]
\begin{tabular}{|c|c|c|c|c|c|c|c|}
\hline
\;Model \;& \;Parametrization \;&\; Free parameters \;\\ \hline
M1 & $q(z)=q_0+q_1z$ & $(H_0,q_0,q_1)$ \\ \hline
M2 & $q(a)=q_0+q_a(1-a)$ & $(H_0,q_0,q_a)$ \\ \hline
M3 & $j(z)=j_0$      & $(H_0,q_0,j_0)$ \\ \hline
M4 & $j(z)=j_0+j_1z$ & $(H_0,q_0,j_0,j_1)$\\ \hline
M5 & $j(a)=j_0+j_a(1-a)$ & $(H_0,q_0,j_0,j_a)$\\ \hline
M6 & $s(z)=s_0$      & $(H_0,q_0,j_0,s_0)$\\ \hline
M7 & $s(z)=s_0+s_1z$ & $(H_0,q_0,j_0,s_0,s_1)$\\ \hline
M8 & $s(a)=s_0+s_a(1-a)$ & $(H_0,q_0,j_0,s_0,s_a)$\\ \hline
\end{tabular}
\caption{Kinematic parametrizations of deceleration parameter, jerk and snap.}
    \label{odels}
\end{table}

\subsection{ Parametrization \texpdf{$q(z)=q_0+q_1z$}{qz=q0+q1*z} }
For a linear parametrization of the deceleration parameter of the form $q=q_0+q_1z$, we can obtain the Hubble parameter from \eqref{Hfromq}:
\begin{align}
    H(z)=H_0(1+z)^{1+q_0-q_1}e^{q_1z}
\end{align}

The dimensionless comoving distance $D_C(z)$ can be obtained from:
\begin{align}
    D_C(z)=\int_0^z\frac{1}{E(z)}dz
\end{align}
where $E(z)\equiv\frac{H(z)}{H_0}$. It can be evaluated to give
\begin{align}
    D_C(z)=e^{q_1}q_1^{q_0-q_1}\left[\gamma(q_1-q_0,q_1)-\gamma(q_1-q_0,q_1(1+z))\right]
\label{Dcq0q1}
\end{align}
where $\gamma$ is incomplete gamma function, defined by
\be
\gamma(a,x)\equiv\int_x^\infty t^{a-1}e^{-t}dt
\ee

The current values (at $z=0$) of jerk, $j_0$, and snap, $s_0$, can be obtained from \eqref{sysqH}:

\be
j_0=q_1+q_0+2q_0^2
\ee
and
\be
s_0=-2q_0-4q_1-7q_0^2-7q_0q_1-6q_0^3
\ee

\subsection{Parametrization \texpdf{$q(z)=q_0+\frac{q_az}{1+z}$}{q(z)=q0+qa*z/(1+z)} }
For a linear parametrization of the deceleration parameter of the form $q=q_0+q_a(1-a) = q_0+\frac{q_az}{1+z}$, we can obtain the Hubble parameter from \eqref{Hfromq}:
\begin{align}
    H=H_0(1+z)^{1+q_0+q_a}\exp\left(-\frac{q_az}{1+z}\right)
\end{align}

The dimensionless comoving distance is
\begin{align}
    D_C(z)=e^{q_a} q_a^{-q_0-q_a} \left[\gamma\left(q_0+q_a,\frac{q_a}{1+z}\right)-\gamma (q_0+q_a,q_a)\right]
\label{Dcq0qa}
\end{align}

The current value of jerk and snap parameters can be obtained from \eqref{sysqH}:
\be
j_0=q_a+q_0+2q_0^2
\ee
and
\be
s_0 = -2q_0 - 2q_a - 7q_0^2 - 7q_0q_a - 6q_0^3\,.
\ee

\subsection{Parametrization \texpdf{$j(z)=j_0$}{j(z)=j0} }

For parametrizations of $j(z)$ the equation for $H(z)$ is quite complicated, coming from (\ref{eq10}). For a constant jerk, $j(z)=j_0$, one can obtain an analytic expression as follows. Eq. \eqref{odefzgeral} reads
\be
(1+z)^2f''(z)-2(1+z)f'(z)+2(1-j_0)f(z)=0
\label{odefzj0}
\ee

If one does the substitution
\be
x\equiv1+z,
\ee
\eqref{odefzj0} reads:
\be
x^2f''(x)-2xf'(x)+2(1-j_0)f(x)=0,
\label{odefxj0}
\ee
which is a Cauchy-Euler equation for $f(x)$. The Cauchy-Euler equation has a well known solution, which can be found by the \textit{ansatz} $f=x^m$. Replacing it on \eqref{odefxj0}, we find:
\be
(m^2-3m+2-2j_0)x^m=0,
\ee
which has solutions
\be
m=\frac{3\pm\sqrt{1+8j_0}}{2}
\ee

So, we have 3 possibilities, according to the sign of $1+8j_0$.
\subsubsection{\texpdf{$1+8j_0>0$}{1+8*j0>0}, or \texpdf{$j_0>-\frac{1}{8}$}{j0>-1/8}}
In this case, let us define the real numbers:
\be
m_1=\frac{3+\sqrt{1+8j_0}}{2},\quad m_2=\frac{3-\sqrt{1+8j_0}}{2}
\ee
and the solution is given by:
\be
f=c_1x^{m_1}+c_2x^{m_2}
\ee
or
\be
H^2=c_1(1+z)^{m_1}+c_2(1+z)^{m_2}
\ee
As $f=H^2$, one has that $f(0)=H_0^2$ and $f'=2HH'$, so $f'(0)=2H_0H'(0)$. As $H'=H\frac{1+q}{1+z}$, $H'(0)=H_0(1+q_0)$ and $f'(0)=2H_0^2(1+q_0)$. So, if one imposes the initial conditions $f(z=0)=H_0^2$, $f'(z=0)=2(1+q_0)H_0^2$, we find
\be
\frac{H^2}{H_0^2}=\left(1+\frac{1+4q_0}{\sqrt{1+8j_0}}\right)\frac{(1+z)^{m_1}}{2} + \left(1-\frac{1+4q_0}{\sqrt{1+8j_0}}\right)\frac{(1+z)^{m_2}}{2}
\ee

\subsubsection{\texpdf{$1+8j_0=0$}{1+8*j0=0}, or \texpdf{$j_0=-\frac{1}{8}$}{j0=-1/8}}
In this case, we have equal solutions $m=m_1=m_2=\frac{3}{2}$, and one has to find the second solution, given the particular solution $f=x^{3/2}$. It is given by $f=x^{3/2}\ln x$, so:
\be
f=c_1x^{3/2}+c_2x^{3/2}\ln x
\ee
or
\be
H^2=(1+z)^{3/2}\left[c_1+c_2\ln(1+z)\right]
\ee
Given the initial conditions $f(z=0)=H_0^2$, $f'(z=0)=2(1+q_0)H_0^2$, we find
\be
\frac{H^2}{H_0^2}=(1+z)^{3/2}\left[1+\left(\frac{1}{2}+2q_0\right)\ln(1+z)\right]
\ee

\subsubsection{\texpdf{$1+8j_0<0$}{1+8*j0<0}, or \texpdf{$j_0<-\frac{1}{8}$}{j0<-1/8}}
In this case, we have complex conjugate solutions
\be
m_1=\alpha+\beta i,\quad m_2=\alpha-\beta i
\ee
where
\be
\alpha=\frac{3}{2},\quad\beta=\frac{1}{2}\sqrt{-1-8j_0}
\ee

So, the solution will be:
\be
f=x^{3/2}\left[c_1\cos(\beta\ln x)+c_2\sin(\beta\ln x)\right]
\ee
or
\be
H^2=(1+z)^{3/2}\left\{c_1\cos[\beta\ln(1+z)]+c_2\sin[\beta\ln(1+z)]\right\}
\ee
Given the initial conditions $f(z=0)=H_0^2$, $f'(z=0)=2(1+q_0)H_0^2$, we find
\be
\frac{H^2}{H_0^2}=(1+z)^{3/2}\left\{\cos[\beta\ln(1+z)]+\frac{1+4q_0}{\sqrt{-1-8j_0}}\sin[\beta\ln(1+z)]\right\}
\ee

The current value of snap parameter can be obtained from \eqref{sysqH}:
\be 
s_0 = -(2+3q_0)j_0.
\ee 
\subsection{Other parametrizations and derived parameters}

For the other parametrizations, namely M4-M8, the expression for $H(z)$ is much more involved and were obtained just numerically. The derived parameters in each case are presented in Table \ref{derived}.

\begin{table}[ht]
\begin{tabular}{|c|c|c|c|}
\hline
Model & Parameterization & $j_0$ & $s_0$ \\ \hline
M1& $q(z)=q_0+q_1z$ & $q_0 + q_1 + 2q_0^2$ & $-2q_0 - 4q_1 - 7q_0^2 - 7q_0q_1 - 6q_0^3$  \\ \hline
M2& $q(z)=q_0+q_a(1-a)$ & $q_0 + q_a + 2q_0^2$ & $-2q_0 - 2q_a - 7q_0^2 - 7q_0q_a - 6q_0^3$    \\ \hline
M3& $j=j_0$ & - & $-(2+3q_0)j_0$ \\ \hline
M4& $j(z)=j_0+j_1z$ & - & $-j_1-(2+3q_0)j_0$   \\ \hline
M5& $j(z)=j_0+j_a(1-a)$ & - & $-j_a-(2+3q_0)j_0$\\ \hline
\end{tabular}
\caption{Derived parameters.}
    \label{derived}
\end{table}


\section{Analyses and Results}

\subsection{Analyses}
In order to constrain the free parameters of kinematic parametrizations, we have used apparent magnitudes from SNe Ia compilation Pantheon \cite{pantheon} and $H(z)$ data compilation from \cite{MaganaEtAl17}.

For SNe Ia, the constraints on free parameters comes from minimizing the $\chi^2$ function:

\be
\chi^2_{SN}=\sum_{i=1}^{1048}\sum_{j=1}^{1048}\left(m_i-m(z_i,\vec{\theta})\right)\left(C^{-1}\right)_{ij}\left(m_j-m(z_j,\vec{\theta})\right)
\ee
where $m_i$ is the observed apparent magnitude of SNe Ia, $\vec{\theta}$ is the parameter vector, $C_{ij}$ is the covariance matrix\footnote{Here, we should mention two facts about the Pantheon data and its covariance matrix. First of all, the Pantheon data are given unsorted in redshift, which is not optimal for the numerical evaluation of the luminosity distance. Thus, we first sort the data in terms of the redshift, then we sort the covariance matrix accordingly. Second, we should mention that we obtained the inverse covariance matrix with the aid of the publicly available Scipy function linalg.inv (\url{https://docs.scipy.org/doc/scipy/reference/generated/scipy.linalg.inv.html}).},  and $m(z_i,\vec{\theta})$ is the predicted magnitude at redshift $z_i$, given by
\be
m(z,\vec{\theta})=5\log_{10}D_L(z,\vec{\theta})+\mathcal{M},
\ee
where $\mathcal{M}$ is a nuisance parameter which encompasses $H_0$ and absolute magnitude $M$ and $D_L(z,\vec{\theta})$ is dimensionless luminosity distance, given, for a spatially flat Universe, as
\be
D_L(z)=(1+z)\int_0^z\frac{dz'}{H(z')/H_0}
\ee

We choose to project over $\mathcal{M}$, which is equivalent to marginalize the likelihood $\like\propto e^{-\chi^2/2}$ over $\mathcal{M}$, up to a normalization constant. This is explained, for instance, in \cite{EscobalEtAl21}.

For $H(z)$ data, {in order to avoid the model dependence of BAO data, we restrict our analysis to the 31 cosmic chronometers data from \cite{MaganaEtAl17}. So,} the $\chi^2$ is given by:
\be
\chi^2_H=\sum_{i=1}^{31}\left[\frac{H_i-H(z_i,\vec{\theta})}{\sigma_{Hi}}\right]^2
\ee
where $H(z_i,\vec{\theta})$ is the predicted Hubble parameter, $H_i$ is the observed Hubble parameter, and $\sigma_{Hi}$ is its uncertainty.

The joint likelihood is then given by
\be
\mathcal{L}=\mathcal{L}_{SN}\mathcal{L}_H
\ee
where
\be
\mathcal{L}_i=Ae^{-\chi_i^2/2}
\ee
is the likelihood for each dataset and $A$ is a normalization constant. The priors we have used for parameters are flat with a large range spanning all the region where the likelihoods are non-negligible, {as can be seen on Table \ref{priors}}. The posteriors are then given by:
\be
p(\vec{\theta})=\pi(\vec{\theta})\mathcal{L}(\vec{\theta})
\ee
where $\pi$ is the prior. In order to probe the posterior distributions, we have used the free open source software emcee \cite{emcee}, which is based on an Affine Parametrization Ensemble Monte Carlo method \cite{affine}.

\begin{table}[t]
\begin{tabular}{|c|c|c|c|c|c|c|}
\hline
\;Model\; & Free parameters & \;$p$\; & \;$\chi^2_{min}$\; & \;$\chi^2_{red}$\; & \;BIC\;     & \;$\Delta$BIC\; \\ \hline
M1    & $(H_0,q_0,q_1)$                & 3   & 1043.80       & 0.97007        & 1064.75 & 0.75        \\ \hline
M2    & $(H_0,q_0,q_a)$                & 3   & 1043.20       & 0.96951        & 1064.15 & 0.15        \\ \hline
M3    & $(H_0,q_0,j_0)$ & 3   & 1043.05       & 0.96937        & 1064.00 & 0       \\ \hline
M4    & $(H_0,q_0,j_0,j_1)$ & 4   & 1043.04       & 0.97027        & 1070.98 & 6.98        \\ \hline
M5    & $(H_0,q_0,j_0,j_a)$ & 4   & 1043.04       & 0.97027        & 1070.98 & 6.98        \\ \hline
M6    & $(H_0,q_0,j_0,s_0)$ & 4   & 1042.96       & 0.97020        & 1070.90 & 6.90        \\ \hline
M7    & $(H_0,q_0,j_0,s_0,s_1)$ & 5   & 1042.69       & 0.97085        & 1077.61 & 13.62       \\ \hline
M8    & $(H_0,q_0,j_0,s_0,s_a)$ &  5   & 1042.82       & 0.97097        & 1077.74 & 13.74       \\ \hline
\end{tabular}
\caption{Kinematic parametrizations of deceleration parameter, jerk and snap, including $\chi^2$ and BIC values.}
    \label{bic}
\end{table}

After generating the Monte Carlo chains, in order to plot the results as contours of marginalized posteriors, we have used the free open source software getdist \cite{getdist}.

\begin{table}[t]
\begin{tabular}{|c|c|c|}
\hline
\;Model\; & Free parameters & Flat Priors \\ \hline
M1    & $(H_0,q_0,q_1)$ & $([50,100],[-5,5],[-20,20])$ \\ \hline
M2    & $(H_0,q_0,q_a)$ & $([50,100],[-5,5],[-20,20])$  \\ \hline
M3    & $(H_0,q_0,j_0)$ & $([50,100],[-5,5],[-10,10])$ \\ \hline
M4    & $(H_0,q_0,j_0,j_1)$ & $([50,100],[-5,5],[-20,20],[-50,50])$ \\ \hline
M5    & $(H_0,q_0,j_0,j_a)$ & $([50,100],[-5,5],[-20,20],[-50,50])$ \\ \hline
M6    & $(H_0,q_0,j_0,s_0)$ & $([50,100],[-5,5],[-20,20],[-50,50])$ \\ \hline
M7    & $(H_0,q_0,j_0,s_0,s_1)$ & $([0,100],[-5,20],[-240,30],[-190,190],[-190,300])$ \\ \hline
M8    & $(H_0,q_0,j_0,s_0,s_a)$ & $([0,100],[-5,20],[-240,30],[-190,190],[-500,500])$\\ \hline
\end{tabular}
\caption{Intervals for the adopted flat priors for each parameter and parametrization.}
    \label{priors}
\end{table}

\subsection{Results}
First of all, in order to do a model comparison among all parametrizations, we have calculated the Bayesian Information Criterion (BIC) \cite{Schwarz78,Liddle04,JesusEtAl16} for them. The BIC is given by:
\be
\text{BIC}=-2\ln\mathcal{L}_{max}+p\ln N=\chi^2_{min}+p\ln N
\ee
where $p$ is the number of free parameters and $N$ is number of data. The BIC values found for all models can be seen on Table \ref{bic}.

\begin{table}[]
\begin{tabular}{|c|c|c|c|c|c|c|c|}
\hline
Model & $H_0$                        & $q_0$                        & $q_1$, $q_a$                        & $j_0$                        & $j_1$, $j_a$                        & $s_0$                        & $s_1$, $s_a$ \\ \hline
M1    & $68.8^{+3.8}_{-3.7}        $ & $-0.51\pm0.12$      & $0.73^{+0.28}_{-0.29}      $                               & $0.75^{+0.41}_{-0.38}      $                               &                              & $-0.24^{+0.54}_{-0.41}     $ &       \\ \hline
M2    & $69.0\pm3.8        $ & $-0.63\pm0.16     $ &                               $1.62^{+0.64}_{-0.66}      $ & $1.79^{+0.89}_{-0.84}      $                              &                              & $4.0^{+3.4}_{-3.0}         $ &       \\ \hline
M3    & $68.8^{+3.7}_{-3.6}        $ & $-0.58\pm0.13     $ &                                                            & $1.15^{+0.56}_{-0.53}      $                              &                              & $-0.25^{+0.40}_{-0.30}     $ &       \\ \hline
M4    & $68.8\pm3.8        $ & $-0.58^{+0.26}_{-0.25}     $                               &                              & $1.2\pm1.9         $ & $0.0^{+3.7}_{-3.6}         $                               & $0.1^{+4.3}_{-3.5}         $ &       \\ \hline
M5    & $68.8\pm3.8        $ & $-0.57^{+0.31}_{-0.30}     $                               &                              & $1.1^{+2.6}_{-2.7}         $ &                               $0.3^{+8.7}_{-8.4}         $ & $0.0^{+9.0}_{-7.9}         $ &       \\ \hline
M6    &   $68.6\pm3.9        $                           &  $-0.53^{+0.32}_{-0.30}     $                                                          &                              &                              $0.6^{+2.5}_{-2.7}         $                                                      &                              & $-2.7^{+6.5}_{-6.9}$      & \\ \hline
M7    & $68.4\pm3.9$ & $-0.45^{+0.52}_{-0.49}$  &                               & $-0.6^{+6.2}_{-6.5}$ &                                                            & $-9^{+34}_{-32}$ & $11^{+55}_{-62}$       \\ \hline
M8    & $68.4^{+4.0}_{-3.9}$ &  $-0.46^{+0.59}_{-0.56}$ &  & $-0.6^{+7.8}_{-8.1}$ & & $-10^{+52}_{-48}$ &  $20^{+100}_{-200}$\\ \hline
\end{tabular}
\caption{Results for all kinematic parametrizations, with mean values of the parameters and 95\% c.l. uncertainties.}
\label{results}
\end{table}


\begin{table}[ht]
\begin{tabular}{|c|c|c|c|c|c|}
\hline
Model &  $H_0$ &  $q_0$  & $j_0$ &  $s_0$ \\ \hline
M1 & $68.8^{+1.9+3.8}_{-1.9-3.7}$ &  $-0.515\pm0.059\pm0.12$      & $0.75^{+0.20+0.41}_{-0.20-0.38}$ & $-0.24^{+0.14+0.54}_{-0.28-0.41}$                                         \\ \hline
M2 & $69.0\pm1.9\pm3.8$ &  $-0.627\pm0.078\pm0.16$      & $1.79^{+0.43+0.89}_{-0.43-0.84}$ & $4.0^{+1.2+3.4}_{-2.0-3.0}$                                        \\ \hline
M3 & $68.8^{+1.9+3.7}_{-1.9-3.6}$ &  $-0.578\pm0.067\pm0.13$ & $1.15^{+0.28+0.56}_{-0.28-0.53}$& $-0.255^{+0.094+0.40}_{-0.21\;\;-0.30}$                                        \\ \hline
\end{tabular}
\caption{Results for kinematic parametrizations M1 - M3, with mean values of $H_0$, $q_0$, $j_0$ and $s_0$ and 68\% and 95\% c.l. uncertainties.}
\label{resultsH0q0j0s0}
\end{table}

As can be seen, models M1 - M3 are favoured by this BIC analysis\footnote{ In order to conclude this, we have used the Jeffreys' scale as explained, for instance, in \cite{KassRaftery95,JesusEtAl16}.}. In Table \ref{results}, we can have a general view of the results for parametrizations M1 - M8. As one can see, the value obtained for $H_0$ is quite similar for all models. Here, we shall mention that we have avoided to use the Hubble constant from local SNe Ia \cite{H0Riess}, as it is currently in tension \cite{H0tension} with high redshift estimates in the context of dynamical models from Cosmic Microwave Background measurements \cite{Planck18}\footnote{We should also mention that one point in the $H(z)$ dataset, at $z=0.47$, from \cite{RatsimbazafyEtAl17}, can have a larger uncertainty than the one estimated by \cite{MaganaEtAl17}. That is why \cite{MaganaEtAl17} {chose to combine the statistical and systematic uncertainties of this point through a midpoint method. That is, they estimate the uncertainty of this point from an arithmetic average of the statistical and systematic uncertainties, instead of using the more recommended method of the quadrature sum of both types of uncertainties.}
However, we have analyzed model M3 with this new uncertainty and the difference was negligible for the parameters in comparison with Table \ref{results}.}. Local SNe Ia data indicate $H_0=73.2\pm1.3$ km/s/Mpc, while CMB data from Planck 2018 indicate $H_0=67.4\pm0.5$ km/s/Mpc. We can see that although our result is independent of dynamics, it is more in agreement with the CMB result. {To be more precise, $H_0$ from M3 is compatible with CMB within 0.7$\sigma$, while it is marginally compatible with SNe Ia data at 1.9$\sigma$.}

The mean values of the deceleration parameter are all compatible, given the uncertainties. The jerk parameter $j_0$ also presents compatibility among all models, at least at 2$\sigma$ c.l. The result for jerk is also in agreement with \cite{hassan2020}, where they have found $j_0=1.038^{+0.061}_{-0.023}$, using BAO+Pantheon SNe Ia+Cosmic Chronometers, in context of model M3. The snap parameter $s_0$, however, presents an incompatibility at the model M2 when compared to models M1 and M3. 

\begin{figure}[ht] 
\begin{center}
\includegraphics[width=\textwidth]{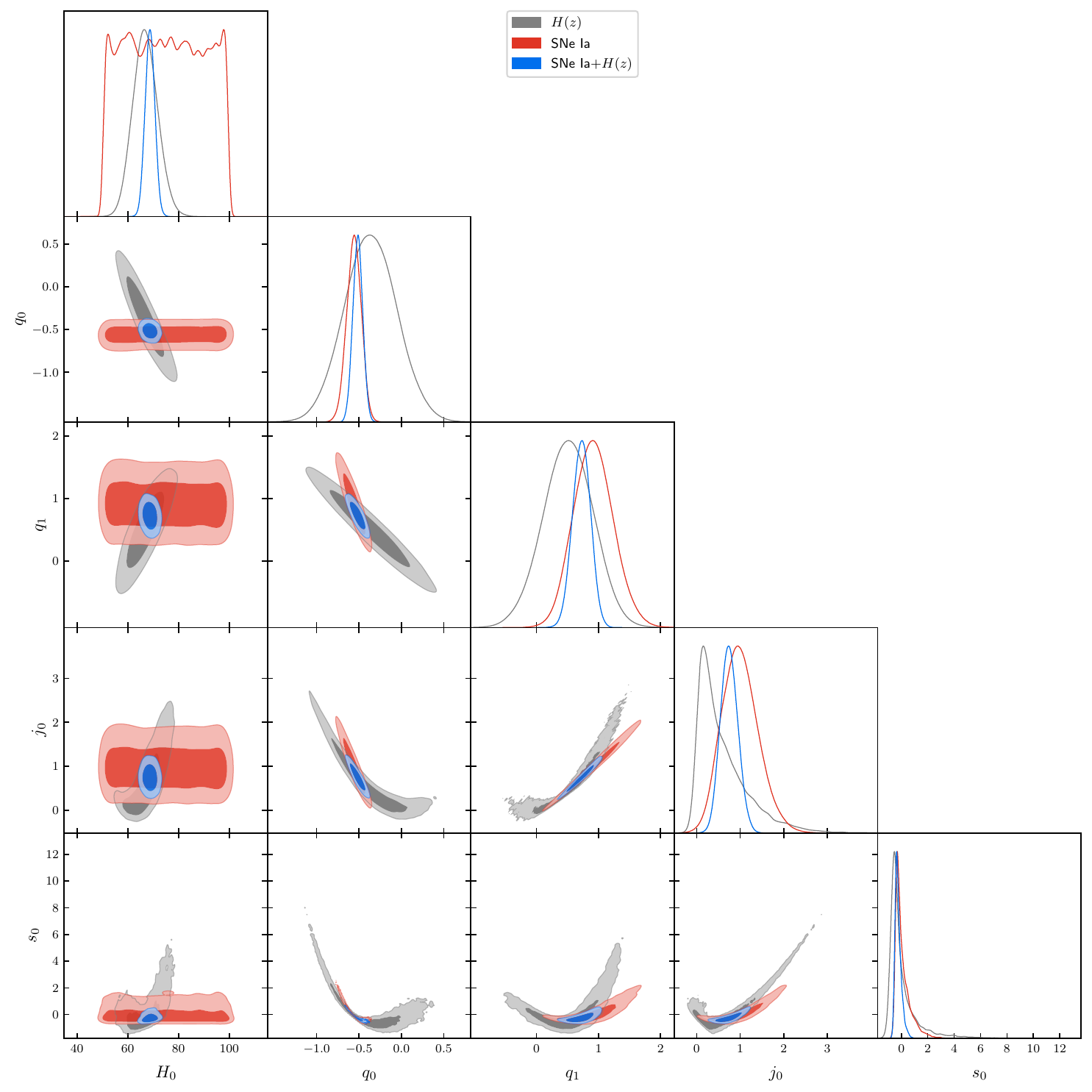}
\end{center}
\caption{\label{qz1} M1: $q(z)=q_0+q_1z$}
\end{figure}

\begin{figure}[ht] 
\begin{center}
\includegraphics[width=\textwidth]{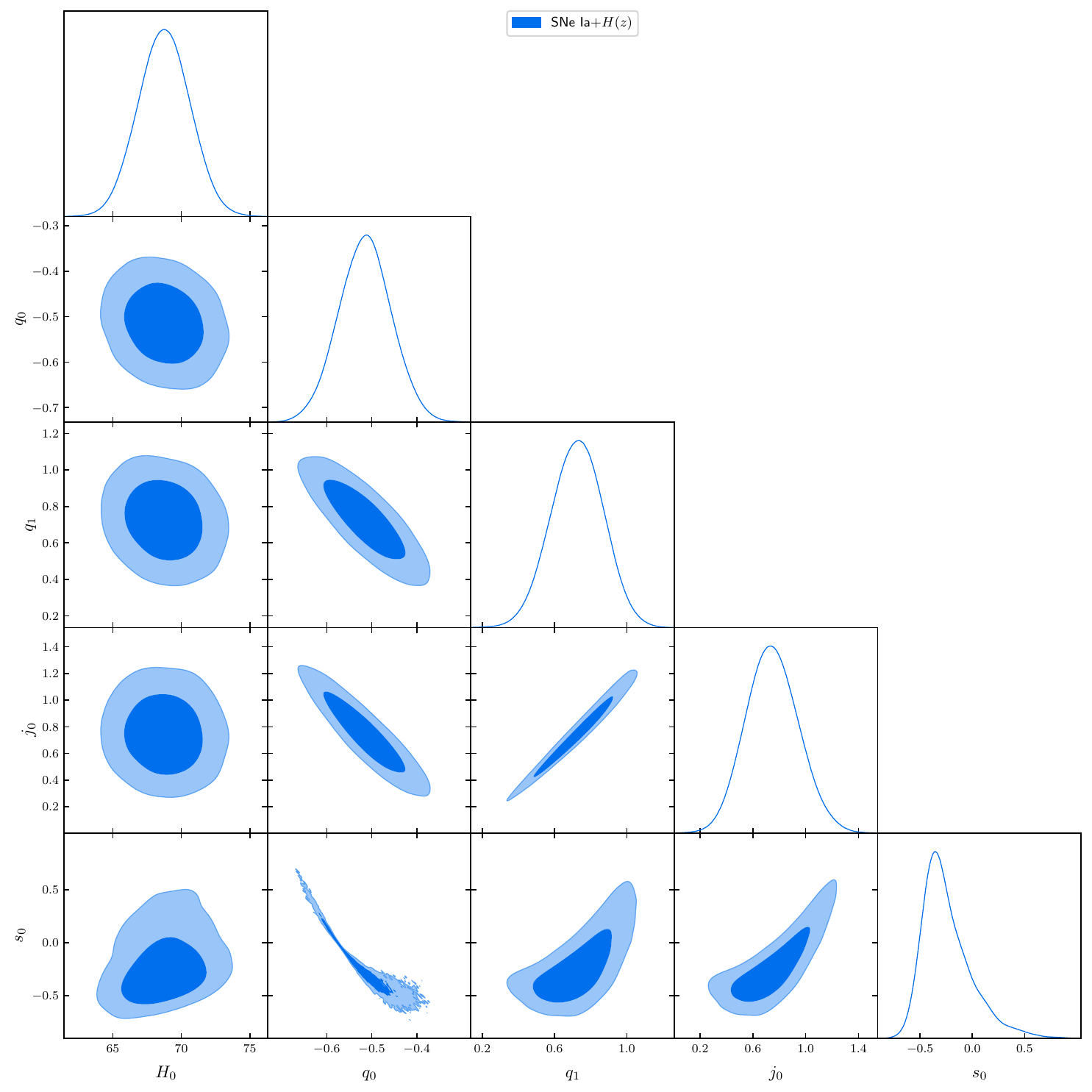}
\end{center}
\caption{\label{qz2} M1: $q(z)=q_0+q_1z$}
\end{figure}

\begin{figure}[ht] 
\begin{center}
\includegraphics[width=\textwidth]{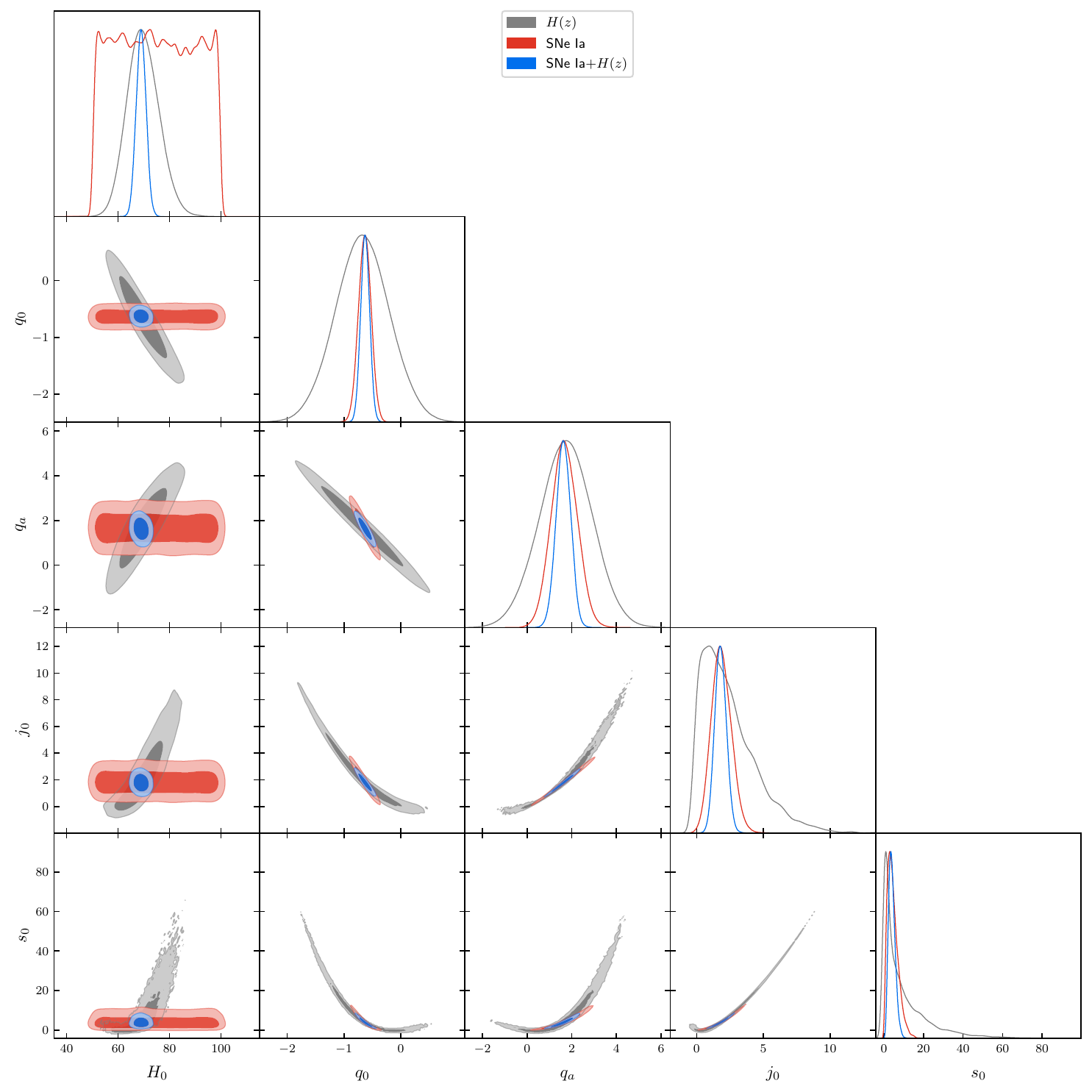}
\end{center}
\caption{\label{qa1} M2: $q(z)=q_0+q_a(1-a)$.}
\end{figure}

\begin{figure}[ht] 
\begin{center}
\includegraphics[width=\textwidth]{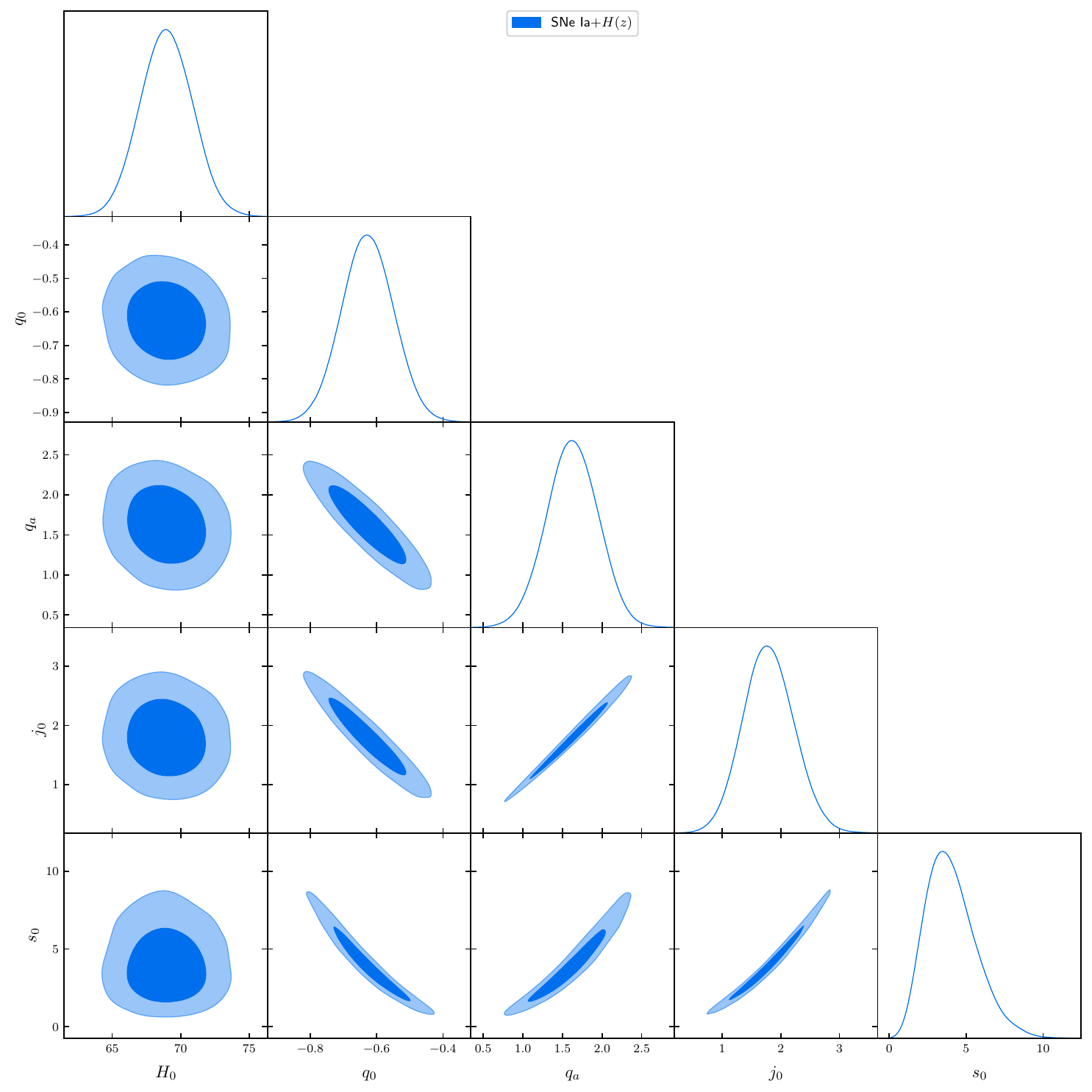}
\end{center}
\caption{M2: $q(z)=q_0+q_a(1-a)$.}
\label{qa2}
\end{figure}

\begin{figure}[ht] 
\begin{center}
\includegraphics[width=\textwidth]{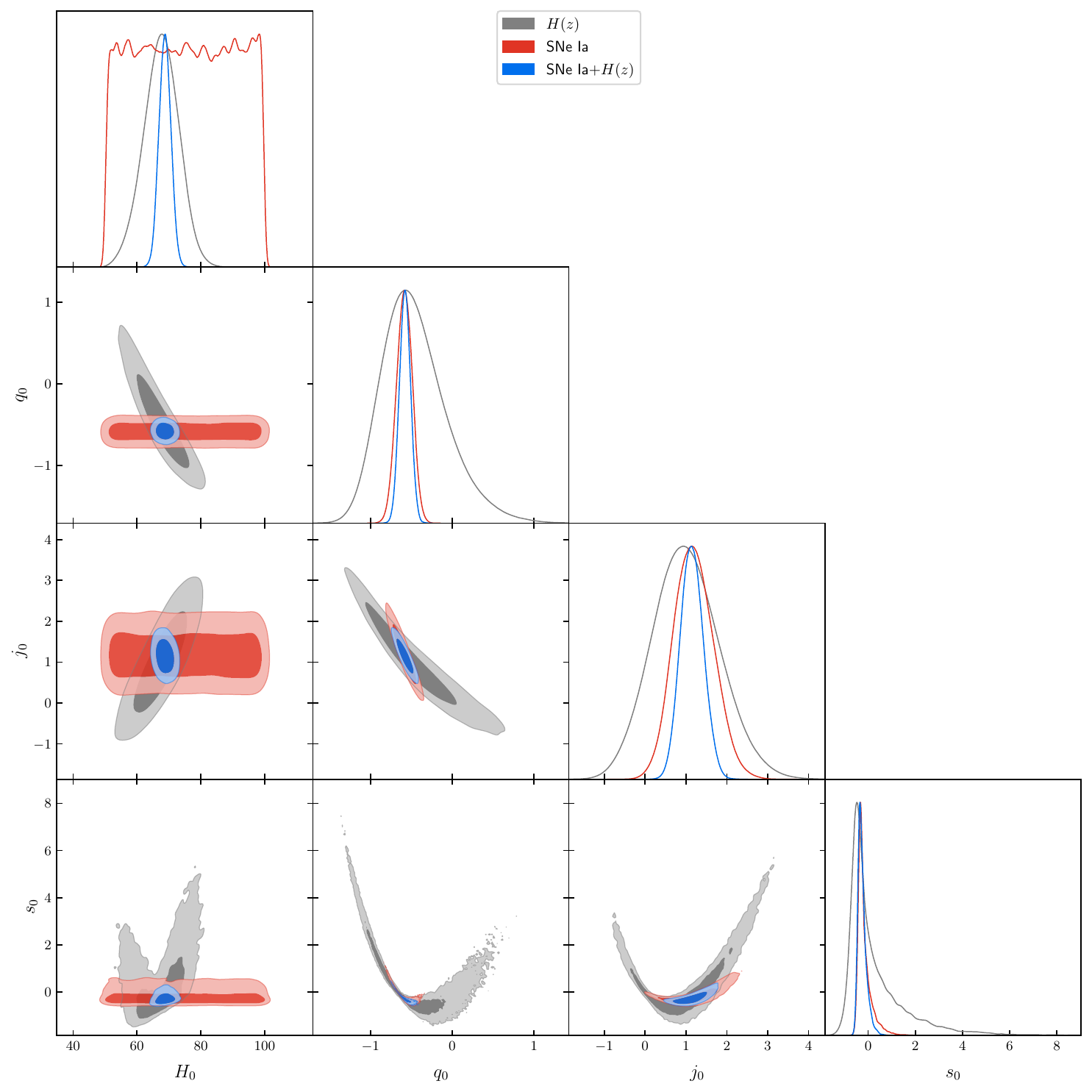}
\end{center}
\caption{\label{j01} M3: $j(z)=j_0$.}
\end{figure}

\begin{figure}[ht] 
\begin{center}
\includegraphics[width=\textwidth]{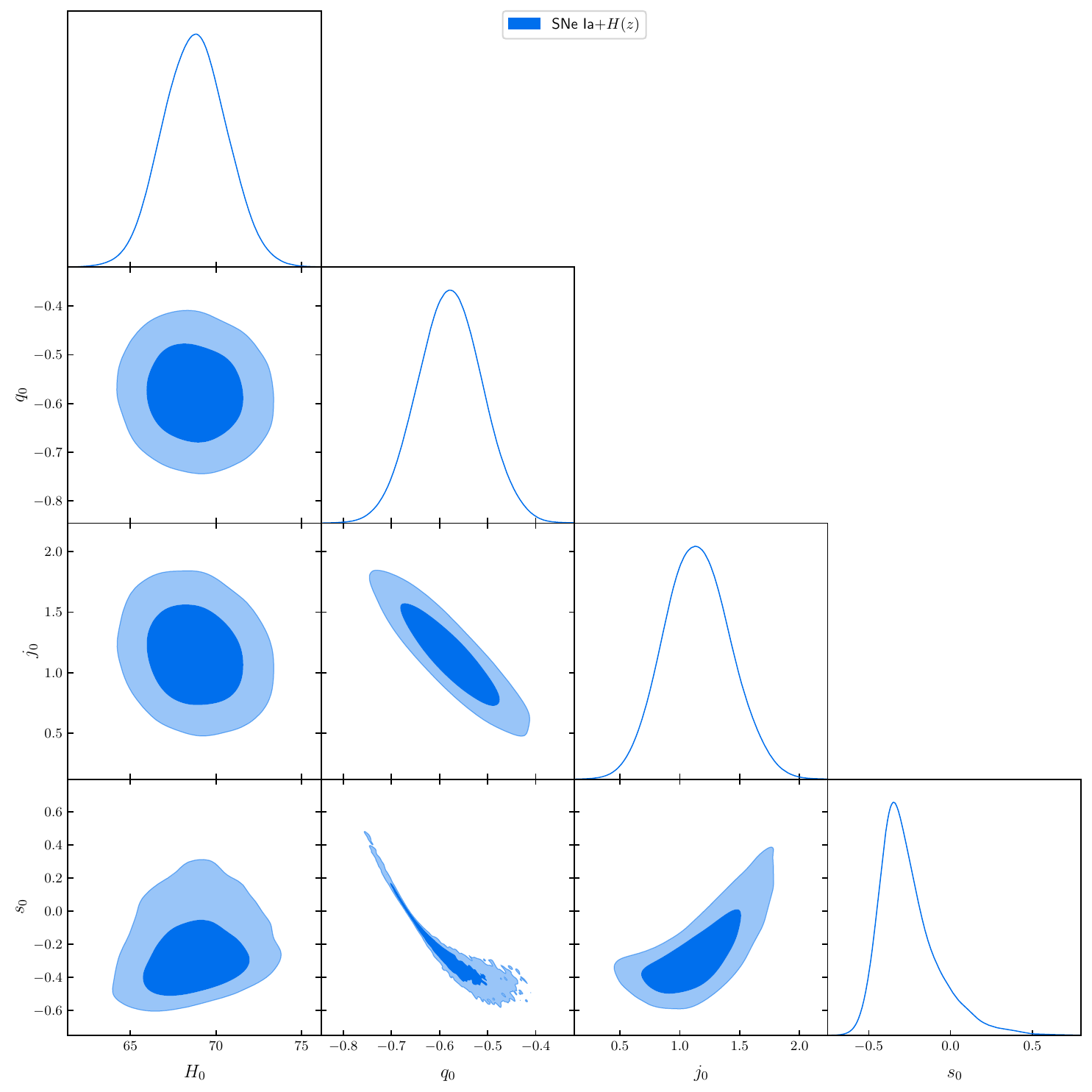}
\end{center}
\caption{M3: $j(z)=j_0$.}
\label{j02}
\end{figure}

As mentioned earlier, models M1 - M3 are favoured by the BIC analysis, so we shall focus on them from now on and present their statistical contours on Figs. \ref{qz1}-\ref{j02}.

As can be seen, on Figs. \ref{qz1}-\ref{j02} we have included both original and derived parameters\footnote{ We should mention that we have obtained the derived parameters constraints directly from the Monte Carlo-Markov chains, as explained, for instance, in th getdist Plot Gallery examples (\url{https://getdist.readthedocs.io/en/latest/plot_gallery.html}).}. We can see the SNe Ia and $H(z)$ data complementarity, mainly for $H_0$, as it is not constrained by SNe Ia. We can also note a big correlation between higher order parameters.

\newpage

\begin{figure}[ht]
    \centering
    \includegraphics[width=\textwidth]{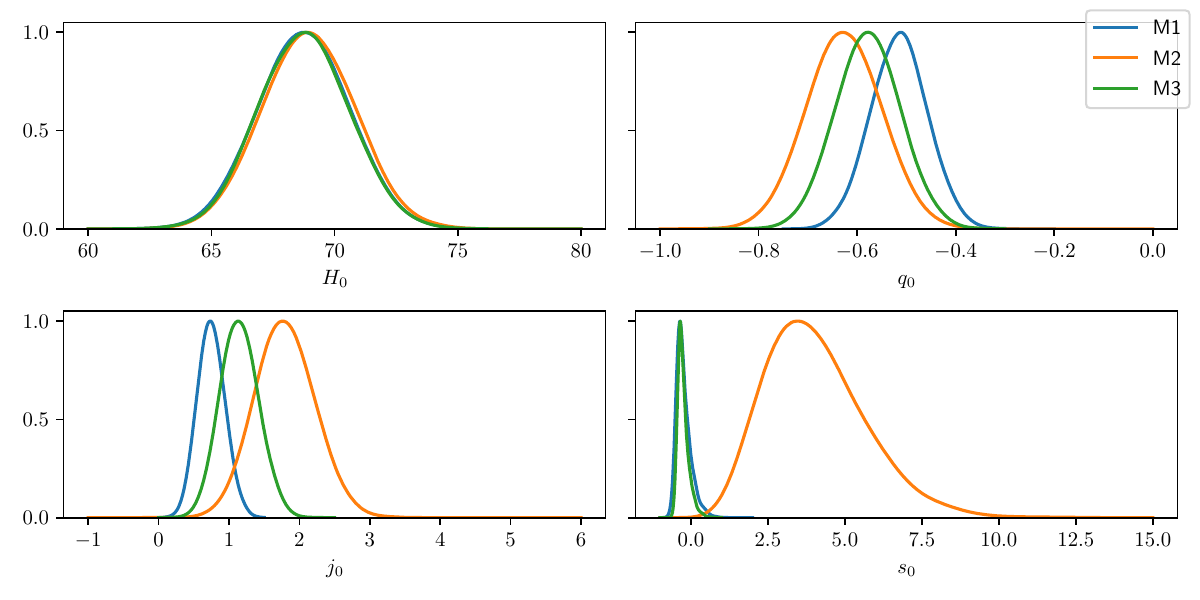}
    \caption{Marginalized posteriors for models M1, M2 and M3.}
    \label{likelihoods}
\end{figure}

One can have a closer look at the models M1, M2 and M3 on Table \ref{resultsH0q0j0s0}, where we show the results for the kinematic parameters for these models, with 1 and 2$\sigma$ c.l. The posteriors for these parameters can be seen on Fig. \ref{likelihoods}.

As one can see from Table \ref{resultsH0q0j0s0}, $H_0$ and $q_0$ are compatible for models M1, M2 and M3; $j_0$ is compatible between models M2 and M3 at 1.25$\sigma$, but is marginally compatible between models M1 and M2 at 2.19$\sigma$. The most drastic result comes from the snap: a 2.65$\sigma$ discrepancy between M2 and M3 and a 2.63$\sigma$ discrepancy between models M2 and M1. We shall mention that this analysis of differences between parameters is approximated mainly for the snap, where we have made a symmetrization of the uncertainties.

As can be seen on Fig. \ref{likelihoods}, the discrepancy for the kinematic parameters is larger for higher derivatives (higher order parameters), so it may arise from different tendencies of the kinematic parameters with redshift. The discrepancy found for the snap is most odd when we compare the models M1 and M2. As one can see on Table \ref{derived}, the expressions for the snap in these models are quite similar. However, the parameters $q_1$ and $q_a$ present a slight difference that is amplified in the terms $7q_0q_1$ and $7q_0q_a$ leading to a huge difference in the snap. Statiscally speaking, one may also note from Figs. \ref{qz2}, \ref{qa2} and \ref{j02} that there is a higher correlation between parameters for the models M1 and M2 where there are more derived parameters ($j_0$ and $s_0$), than model M3, where there is less derived parameters ($s_0$ only). The introduction of more derived parameters may induce an artificial correlation between parameters that can lead to discrepancies like the one found for the snap.

\section{Conclusions}
We have analyzed 8 kinematical cosmological parametrizations against $H(z)$ and SNe Ia data. A Bayesian comparison favoured models where the deceleration parameter is linearly dependent on the redshift (M1), on the scale factor (M2) and a model where the jerk is constant (M3). These models were compatible concerning the Hubble constant and the current deceleration parameter, were marginally compatible in the current value of the jerk and were incompatible in the current value of the snap.

According to the series convergence argument, model M2 should be favoured, as it is an expansion on the scale factor. However, if we believe that the flat $\Lambda$CDM model is the correct underlying dynamical model, we should expect that model M3 should be favoured, as it has the flat $\Lambda$CDM model as an special case (with $j_0=1$). Also, M3 is slightly favoured by the BIC analysis. The current data, however, are not enough to discard any of these three models {(M1, M2 and M3)}.


In order to determine which value is correct for the snap, more research is needed, which may include but not necessarily restricted to: (i) more data can be used to distinguish among the three models and to discard one of them; (ii) other methods which are independent from the dynamics can be used, like non-parametric methods, like Gaussian Processes, Principal Component Analysis (PCA), Neural Networks etc.

Some of these possibilities can be explored in a forthcoming issue.

\begin{acknowledgments}
SHP acknowledges financial support from  {Conselho Nacional de Desenvolvimento Cient\'ifico e Tecnol\'ogico} (CNPq)  (No. 303583/2018-5 and 308469/2021-6). This study was financed in part by the Coordena\c{c}\~ao de Aperfei\c{c}oamento de Pessoal de N\'ivel Superior - Brasil (CAPES) - Finance Code 001.
\end{acknowledgments}


\end{document}